\documentstyle[twocolumn,aps,prl,psfig]{revtex}
\catcode`\@=11
\def\maketitle2{\par 
\begingroup
\let\cite\@bylinecite
\def\thefootnote{\fnsymbol{footnote}}%
\twocolumn[\@maketitle2\vskip2pc]%
\thispagestyle{plain}\@thanks
\endgroup
\def\thefootnote{\arabic{footnote}}%
\setcounter{footnote}{0}%
\let\maketitle2\relax \let\@maketitle2\relax
\let\@thanks\relax \let\@authoraddress\relax \let\@title\relax
\let\@date\relax \let\thanks\relax \let\@abstract\relax
\let\@pacs\relax}
\def\abstract#1{\gdef\@abstract{{\par 
\bgroup
\ifdim\prevdepth=-1000pt \prevdepth0pt\fi
\hsize\columnwidth
\dimen0=-\prevdepth \advance\dimen0 by17.5pt \nointerlineskip
\small\vrule width 0pt height\dimen0 \relax}{~~}#1\egroup}}
\def\pacs#1{\gdef\@pacs{{\par 
\bgroup
\hsize\columnwidth \parindent0pt
\ifdim\prevdepth=-1000pt \prevdepth0pt\fi
\dimen0=-\prevdepth \advance\dimen0 by20pt\nointerlineskip
\egroup} PACS numbers:~#1}}
\def\@maketitle2{
\@preprint
\@title
\ifdim\prevdepth=-1000pt \prevdepth0pt\fi
\@authoraddress
\@date
\begin{list}{}{\leftmargin=0.10753\textwidth \rightmargin=\leftmargin
\itemsep=1pc\partopsep=-1pc}
\item\@abstract
\item\@pacs
\end{list}
}
\catcode`\@=12
\begin{document}
\draft
\title{Self-similar Magnetoresistance of Fibonacci Ultra-Thin Magnetic
Films}

\author{C.  G.  Bezerra$^{1}$, J.  M.  de Ara\'ujo$^{1,2}$, C.  Chesman$^{1}$,
and E.  L.  Albuquerque$^{1}$}

\address{$^1$Departamento de F\'\i sica Te\'orica e Experimental, Universidade
Federal do Rio Grande do Norte, \\59072-970, Natal-RN, Brazil \\$^2$Departamento
de Ci\^encias Naturais, Universidade Estadual do Rio Grande do Norte,
\\59610-210, Mossor\'o-RN, Brazil}

\date{\today}

\abstract {\small{ We study numerically the magnetic properties (magnetization
and magnetoresistance) of ultra-thin magnetic films (Fe/Cr) grown following the
Fibonacci sequence.  We use a phenomenological model which includes Zeeman,
cubic anisotropy, bilinear and biquadratic exchange energies.  Our physical
parameters are based on experimental data recently reported, which contain
biquadratic exchange coupling with magnitude comparable to the bilinear exchange
coupling. When biquadratic exchange coupling is sufficiently large a striking
self-similar pattern emerges.  }}

 \maketitle2 \narrowtext

The discovery of quasicrystals in 1984 \cite{1} aroused a great interest, both
theoretically and experimentally, in quasiperiodic systems.  One of the most
important reason for that is because they can be defined as an intermediate
state between an {\it ordered} crystal (their definition and construction follow
purely deterministic rules) and a {\it disordered} solid (many of their physical
properties exhibit an erratic-like appearance) \cite{2}.  On the theoretical
side, a wide variety of particles, namely electrons \cite{3}, phonons\cite{4},
plasmon-polaritons \cite{5}, spin waves \cite{6}, etc, have been studied. 
A quite complex {\it fractal energy spectrum}, which
can be considered as their basic signature, is a common feature of these
systems.  On the experimental side, the procedure to grow quasiperiodic
superlattices became standard after Merlin {\it et al} \cite{7}, who reported
the realization of the first quasiperiodic superlattice following the Fibonacci
sequence by means of molecular beam epitaxy (MBE).

Parallel to these developments in the field of quasicrystals, the properties of
magnetic exchange interactions between ferromagnetic films separated by
non-magnetic spacers have been also widely investigated\cite{8}.  The discovery
of physical properties like antiferromagnetic exchange coupling \cite{9}, giant
magnetoresistance (GMR) \cite{10}, oscillatory behavior of the exchange coupling
\cite{11}, and biquadratic exchange coupling (BEC) \cite{12}, made these films
excellent options for technological applications and attractive objects of
research.  For example, GMR in magnetic films has been widely considered for
applications in information storage technology \cite{13}.

It is known that GMR also occurs in non-periodic granular systems, like Cu-Co
alloys consisting of ultrafine Co-rich precipitate particles in Cu-rich matrix
\cite{14}.  Due to the fact that precipitate particles of these heterogeneous
alloys have an average diameter and an average spacing similar to magnetic
films, the origin of GMR in granular systems is also similar to the one found in
magnetic films \cite{15}.  Therefore, quasiperiodic systems which present
magnetoresistive properties can be a first step for a better understanding of
magnetoresistance in granular systems.  On the other hand, from a technological
point of view (as we will show later in this letter) the BEC associated with
quasiperiodicity permit us to control magnetic field regions, where
magnetoresistance remains almost constant before saturation.

The aim of this work is to investigate the influence of quasiperiodicity on the
magnetic properties of ultra-thin magnetic films.  In particular, we are
interested in Fe/Cr(100) structures, which follow a Fibonacci sequence, whose
experimental magnetic parameters were recently reported by Rezende {\it et al}
\cite{16}.

A Fibonacci structure can be grown experimentally by juxtaposing two building
blocks $A$ and $B$ following a Fibonacci sequence.  In our specific case we
choose Fe as the building block $A$ and Cr as the building block $B$.  A
Fibonacci sequence $S_{N}$ is generated by appending the $N-2$ sequence to the
$N-1$ one, {\it i.e.}, $S_{N}=S_{N-1}S_{N-2}$ ($N \geq 2$).  This construction
algorithm requires initial conditions which are chosen to be $S_{0}=B$ and
$S_{1}=A$.  The Fibonacci generations are:  $S_0=\left[ B \right]$, $S_1=\left[
A \right]$, $S_2=\left[A B \right]$, $S_3=\left[ AB A \right]$, etc.  Therefore,
the well known trilayer Fe/Cr/Fe is the magnetic counterpart of the third
Fibonacci generation ($A/B/A$).  We remark that only odd Fibonacci generations
have a magnetic counterpart, because they start and finish with an $A$ (Fe)
building block.  Fig.\ 1 shows schematically the third and fifth Fibonacci
generations and their magnetic counterpart, where $t$ ($d$) is the thickness of
a single Fe layer (single Cr layer).  It is important to note a double Fe layer
whose thickness is $2t$ in the fifth generation corresponding to a double letter
$A$.  It is easy to show that the quasiperiodic magnetic films, for any
Fibonacci generation, will be composed by single Cr layers, single Fe layers and
double Fe layers.

We consider the ferromagnetic films with magnetization in the plane $xy$ and
take the $z$ axis as the growth direction (see Fig.\ 1).  The very strong
demagnetization field, generates by tipping the magnetization out of plane, will
suppress any tendency for the magnetization to tilt out of plane.  The global
behavior of the system is well described by a simple theory in terms of the
magnetic energy per unit area \cite{16}, {\it i.e.},

\begin{equation} $$E_{T}=E_{Z}+E_{bl}+E_{bq}+E_{a},$$ \end{equation}

\noindent where $E_{Z}$ is the Zeeman energy, $E_{bl}$ is the bilinear energy,
$E_{bq}$ is the biquadratic energy and $E_{a}$ is the cubic anisotropy energy.
It is usual to write the total magnetic energy in terms of experimental
parameters (or effective fields) of each interaction,

\begin{eqnarray} \lefteqn {{{E_{T}} \over
{tM_{S}}}=\sum_{i=1}^n(t_{i}/t){\{-H_{0}\cos(\theta_{i}-\theta_{H})+ {1 \over
8}H_{a}{{\sin}^2{(2 \theta_{i})}}\}}} \nonumber \\ &&
+\sum_{i=1}^{n-1}{\{-H_{bl}\cos(\theta_{i}-\theta_{i+1}) +H_{bq}{{\cos}^2{
(\theta_{i}-\theta_{i+1})}}\}}, \end{eqnarray}

\noindent where $t$ is the thickness of a single Fe layer and we assume
$M_{i}=M_{S}$.  $H_{bl}$ is the conventional bilinear exchange coupling field
which favors antiferromagnetic alignment (ferromagnetic alignment) if negative
(positive).  We are concerned here to the case $H_{bl} < 0$ because
magnetoresistive effects occur only for this case.  $H_{bq}$ is the BEC field,
which is responsible for a $90^{\circ}$ alignment between two adjacent
magnetizations and is experimentally found to be positive \cite{12}.  $H_{a}$ is
the cubic anisotropy field which renders the $(100)$ direction an easy
direction.  $H_{0}$ is the external in-plane magnetic field and $\theta_{H}$ is
its angular orientation.  From now on we consider $\theta_{H}=0$, which means
that the magnetic field is applied along the easy axis.  The thickness and the
angular orientation of the {\it i-th} Fe layer are given by $t_{i}$ and
$\theta_{i}$, respectively.

The equilibrium positions of the magnetizations $\{ \theta_{i} \}$ are
numerically calculated by minimizing the magnetic energy given by (2).  It
should be remarked that it has proved difficult for us to generate accurately
configurations for larger structures, mainly when the BEC is strong \cite{17}.
However, we got results in sufficiently large generations to infer important
informations about the effect of the quasiperiodicity.

Theoretically, the spin-dependent scattering is accepted as responsible for the
GMR effect \cite{15}.  It has been shown that GMR varies linearly with
$\cos(\Delta \theta )$ when electrons form a free-electrons gas, {\it i.e.},
there is no barriers between adjacent films \cite{18}.  Here, $\Delta \theta$ is
the angular difference between adjacent magnetizations.  In metalic systems like
Fe/Cr this angular dependence is valid and once the set $\{ \theta_{i} \}$ is
found, we obtain normalized values for magnetoresistance, {\it i.e},

\begin{equation} R(H_{0}) /
 R(0)=\sum_{i=1}^{n-1}{[{1-\cos({\theta_{i}-\theta_{i+1}})]} / {2(n-1)}},
 \end{equation} \noindent where $R(0)$ is the resistance at zero field.

Now we present numerical calculations for the magnetization and the
magnetoresistance curves for Fibonacci ultra-thin magnetic films. The physical
motivation for that is because the Fibonacci quasiperiodic structure can exhibit
magnetic properties not found in the periodic case \cite{6}, and the long range
correlations induced by the construction of this system are expected to be
reflected someway in the magnetoresistance curves. We have considered physical
parameters based on realistic values of the magnetic coupling fields, whose
whose experimental data were recently reported \cite{16}. We assume the
cubic anisotropy field $H_{a}=0.5$ kOe, corresponding to Fe(100) with $t>30$
\AA\ growth by sputtering \cite{8}.  We choose the bilinear and the biquadratic
fields, $H_{bl}$ and $H_{bq}$, such that their values lie in three regions of
interest:  i) close to the region of first antiferromagnetic-ferromagnetic
transition where $H_{bl}$ is moderate \cite{12}; ii) near to the maximum of
first antiferromagnetic peak, where $H_{bl}$ reach its maximum value \cite{10};
and iii) in the second antiferromagnetic peak, where $H_{bl}$ is small and equal
to $H_{bq}$ \cite{19}.

In Fig.\ 2 we show the curves of the normalized magnetization and
magnetoresistance versus the magnetic field, for the third Fibonacci generation
(corresponding to the Fe/Cr/Fe trilayer).  We assumed $H_{bl}=-0.15$ kOe and
$H_{bq}=0.05$ kOe ($|H_{bl}|>H_{bq}$).  These parameters correspond to a
realistic sample with Cr thickness equal to $15$ \AA.  From there one can
identify two first order phase transitions at $H_{1} \sim 100$ Oe and $H_{2}
\sim 220$ Oe.  Also there are three magnetic phases presented:  i) an
antiferromagnetic phase ($H_{0}<100$ Oe); ii) a $90^{\circ}$ phase ($100$ Oe
$<H_{0}<220$ Oe); iii) and a saturated phase ($H_{0}>220$ Oe).  We remark that
our numerical calculations indicate that a first order phase transition occurs
when $H_{a} > 2(|H_{bl}|+2H_{bq})$.  Since the transition magnetic fields are
the same for both the magnetization and the magnetoresistance, from now on we
concentrate our discussion on the magnetoresistance curves, because it is easier
to investigate their self-similar pattern.

Fig.\ 3 shows the normalized magnetoresistance curves for the fifth(a) and
seventh(b) Fibonacci generations with the same experimental parameters
considered in Fig.\ 2.  In Fig.\ 3(a) we can identify four first order phase
transitions, where each one is due to a $90^{\circ} $ jump of magnetization.
This behavior is always displayed when the BEC is present in the magnetic energy.
Previous works on phase diagrams have looked carefully the origin and features
of the so-called $90^{\circ}$ phase \cite{17,19}.
For the seventh generation there are eight first order phase transitions and
nine magnetic phases are present from the antiferromagnetic phase ($H_{0} < 38$
Oe) to the saturated one ($H_{0} > 440$ Oe).  Note a clear {\it self-similar
pattern} of magnetoresistance curves by comparing Fig.\ 2 and Fig.\ 3, {\it
i.e.}, the pattern of the trilayer Fe/Cr/Fe is always present in the next
generations.  On the contrary, when $H_{bl}=-1.0$ kOe and $H_{bq}=0.1$ kOe
($|H_{bl}|>>H_{bq}$), which correspond to a sample with Cr thickness equal to 10
\AA, {\it the self-similarity is not observed}, as it is shown in Fig.\ 4.  For
this set of parameters, the majority of phase transitions are of second order
and we have found numerically that this occurs when $H_{a}<2(|H_{bl}|+2H_{bq})$.
However, when the ratio between $H_{bq}$ and $H_{bl}$ is increased ($|H_{bl}|=H_{bq}=35$ Oe), we observe again
a striking {\it self-similar pattern} (see Fig.\ 5), where each new transition
occurs for a value of magnetic field which is about a half of the previous one.
For this set of parameters the magnetoresistance is approximately 1/2 its value
at zero magnetic field, because the magnetizations of the adjacent Fe films are
nearly perpendicular to each other due to the strong biquadratic field.  For the
third generation, Fig.\ 5(a), there is only one transition at $H_{1} \sim 70$ Oe
and two magnetic phases:  a $90^{\circ}$ phase at $H_{0} < 70$ Oe and a
saturated phase at $H_{0} > 70$ Oe \cite{19}.  In the fifth generation, Fig.\
5(b), there are two transitions at $H_{1} \sim 70$ Oe and $H_{2} \sim 140$ Oe,
respectively.  For the seventh generation, as one can see from Fig.\ 5(c), there
are three transitions at $H_{1} \sim 35$ Oe, $H_{2} \sim 70$ Oe and $H_{3} \sim
140$ Oe.

From the numerical results above discussed, we can infer that the magnetoresistance
exhibits a self-similar behavior when: (a) $H_{bq}$ is comparable to $H_{bl}$,
and (b) there is a first order phase transition (see Fig.\ 3 and Fig.\ 5).
A possible explanation for that is because the BEC reinforces the quasiperiodic
order, which is responsible by the self-similarity
in quasiperiodic systems.  This is an unexpected effect of this unusual exchange
coupling and, as far as we know, this is the first system which presents
magnetoresistance with self-similar properties. Besides, from a technological point of
view, magnetoresistance with almost constant regions (Fig.\ 3 and Fig.\ 5) opens
new perspectives in information storage technology by the possibility of a
recording system with more than two states.  Certainly Fibonacci ultra-thin
magnetic films can be realized experimentally following the procedures of
ref.[12] or [20] to grow the samples.

We would like to thank A.  M.  Mariz, N.  S.  Almeida and G. M. Viswanathan for
fruitful discussions, and the Brazilian Research Council CNPq for partial financial
support.  We are also grateful to A.  Albino Jr. for the Fig.\ 1 and CESUP-RS
where part of numerical calculations was done.

\begin{figure} \centerline{\psfig{figure=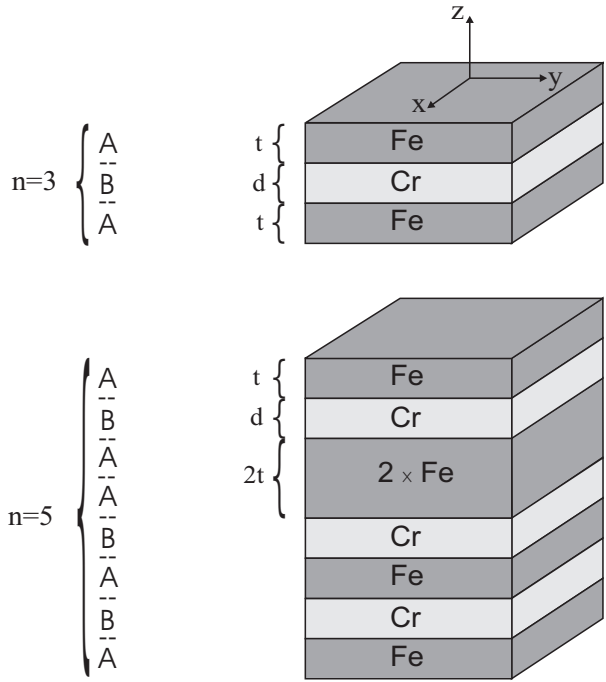,height=9cm,width=8cm}}
\bigskip \caption{The third and fifth Fibonacci generations and their magnetic
counterpart.}

\end{figure}

\begin{figure}

\centerline{\psfig{figure=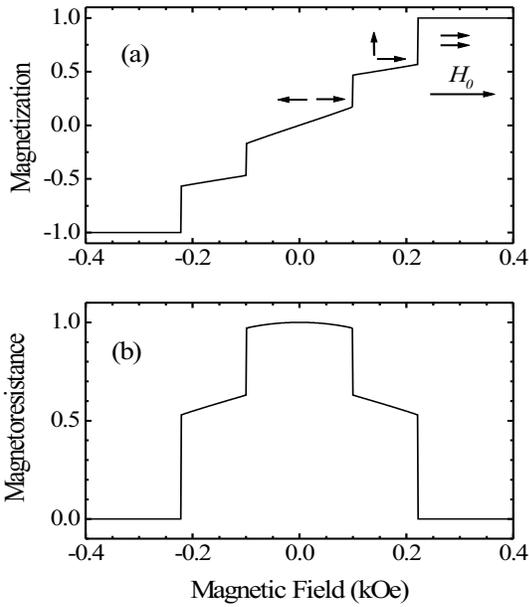,height=9cm,width=8cm}} \caption{Magnetization
(a) and magnetoresistance (b) versus magnetic field for the third Fibonacci
generation with $H_{bl}=-150$ Oe and $H_{bq}=50$ Oe.  In Fig.\ 2(a) the arrows
indicate the relative positions of the magnetizations in each phase.}

\end{figure}

\begin{figure} \centerline{\psfig{figure=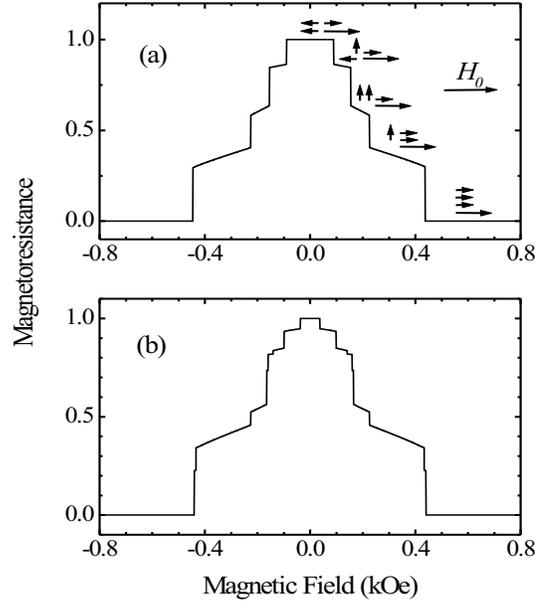,height=9cm,width=8cm}}

\caption{Magnetoresistance for the fifth (a) and seventh (b) Fibonacci
generations with the same parameters of Fig.\ 2.  In Fig.\ 3(a) the relative
positions of magnetizations are indicated by the arrows, and the Fe double layer
is indicated by the bigger arrow.}

\end{figure}

\begin{figure}

\centerline{\psfig{figure=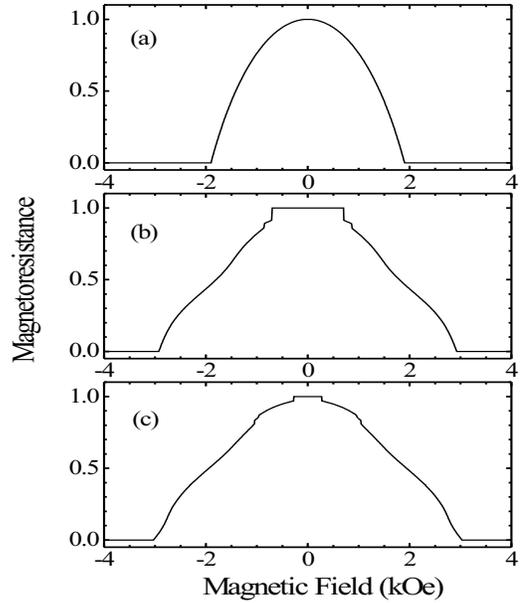,height=9cm,width=8cm}}
\caption{Magnetoresistance for the third (a), fifth (b) and seventh (c)
Fibonacci generations with $H_{bl}=-1.0$ kOe and $H_{bq}=0.1$ kOe.}

\end{figure}

\begin{figure}

\centerline{\psfig{figure=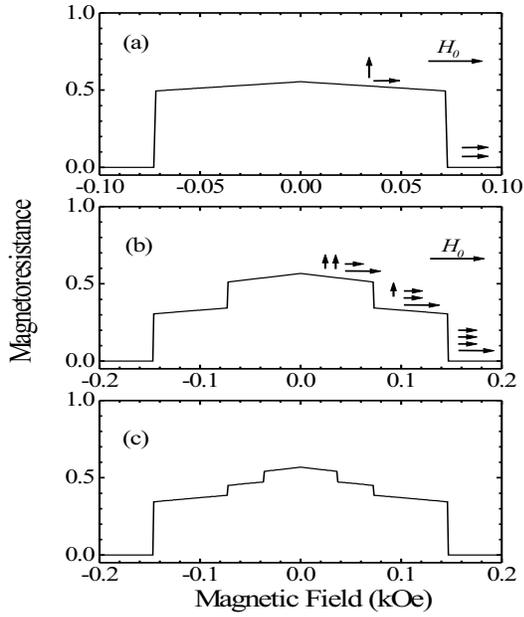,height=9cm,width=8cm}}
\caption{Magnetoresistance for the third (a), fifth (b) and seventh (c)
Fibonacci generations with $|H_{bl}|=H_{bq}=35$ Oe, which correspond to a sample
with Cr thickness equal to $25$ \AA.  Note a striking self-similar pattern.  In
Fig.\ 5(a) and Fig.\ 5(b) the arrows indicate the relative positions of the
magnetizations in each phase.}  \end{figure}

\end{document}